\documentstyle[aps,epsfig,float,twocolumn,prl]{revtex}
\newcommand{\be}{\begin{equation}}
\newcommand{\ee}{\end{equation}}
\begin{document}
\twocolumn[\hsize\textwidth\columnwidth\hsize\csname @twocolumnfalse\endcsname
\draft
\title{Quantum Dynamics in Non-equilibrium Strongly Correlated Environments}
\author{M. B. Hastings$^{1,2}$, I. Martin$^2$, and D. Mozyrsky$^{1,2}$}
\address{
$^1$Center for Nonlinear Studies and $^2$Theoretical Division, Los
Alamos National Laboratory, Los Alamos, NM 87545,
hastings@cnls.lanl.gov }
\date{June 28, 2002}
\maketitle
\begin{abstract}
We consider a quantum point contact between two Luttinger liquids
coupled to a mechanical system (oscillator).
For non-vanishing bias, we find an effective oscillator temperature
that depends on the Luttinger parameter.  A generalized
fluctuation-dissipation relation connects the decoherence and
dissipation of the oscillator to the current-voltage
characteristics of the device.  Via a spectral representation, this
result is generalized to arbitrary leads in a weak tunneling regime.
\vskip2mm
\end{abstract}

] The quest to build a scalable quantum computer has recently lead to a series
of spectacular experiments where macroscopic quantum states were coherently
manipulated and measured\cite{naka,saclay,kansas}.  These experiments for the
first time give an opportunity to study the effects of indirect continuous
measurement
\cite{mandelstam} on an individual quantum system.   Understanding the
measurement process is not only important for the development of a quantum
computer, but also is a fundamental problem of quantum mechanics.  In some
cases, theoretical analysis of measurement process based on explicit models
describing the coupling between the system and electrical measurement apparatus
has been performed
\cite{gurvitz,shnirRMP,korotkov,goran,dima}.

Correlations play a crucial role in protecting quantum coherence in macroscopic
systems and enabling manipulation of the quantum states. In sub-micrometer
electronic systems, the Coulomb interaction becomes important and can lead to
Coulomb blockade,
a subject of intensive research both in the contexts of ``classical''
and quantum-coherent electronic devices.  The more subtle effects of {\it itinerant}
electron-electron interactions, however, have not been studied in the context
of quantum measurement.  These effects are often important since in order to
interface with a quantum device, at least a part of the apparatus has to be
scaled down to the device size.  The study of such correlation effects is therefore
important in developing the understanding of a realistic quantum measurement.

To analyze the role of correlations in the measurement apparatus, we study here a specific example
of two Luttinger liquid leads electrically coupled to a quantum system, with the tunneling current
being influenced by a coordinate of the system.  The Luttinger liquids are a generalization of the
Fermi liquids to one dimension, where electron-electron interactions lead to dramatic
renormalization effects near the Fermi surface\cite{lutt}.   The interactions in the Luttinger
liquids can be parametrized by the dimensionless constant $g$, which describes repulsive
interactions if $g <1$, attractive if $g > 1$, and a non-interacting Fermi liquid for $g = 1$. One
experimental realization of Luttinger liquids is carbon nanotubes\cite{dekker}.
We develop the formalism for an arbitrary quantum system, which can be, for instance, a two-level
system (qubit), or a quantum oscillator.  For $g>1$, we find that the effect
of the measurement is equivalent  to coupling to a heat bath, with an
effective temperature reduced relative to the Fermi liquid case\cite{dima}.
The effective temperature depends on the density of states of the Luttinger
liquids at the tunnel contact, $T_{\rm eff}=(\alpha+1)^{-1}qV/2$, where the current $I\propto
V^{\alpha+1}$.  The exponent $\alpha$ depends on the tunneling geometry and on the Luttinger
parameter $g$.  This provides an interesting example in which the Luttinger parameter determines,
not an exponent, but a prefactor for a universal expression which is of first order in the bias
voltage.  It had been hoped that the two-terminal conductance of a one-dimensional wire would be
proportional to $g$\cite{2c}, providing another such example, but this turns out not to be the
case\cite{2cw}.  Via a spectral representation, we find a relation valid for any leads in the
weak tunneling regime with tunneling particles of charge $q$
\be \label{dvdi} T_{\rm eff}=(qI/2)({\rm d}V/{\rm d}I). \ee

Examples of experimental realizations of our model as applied to
the measurement of a quantum oscillator are shown in
Fig.~\ref{fig:setup}.  In Fig.~1(a) and Fig.~1(b), the tunnel
junction is formed by a nanotube (Luttinger liquid) and a metal
(Fermi liquid), while in Fig.~1(c) and 1(d) both sides of the
junction are Luttinger liquids. In the first example, Fig.~1(a),
the tunnel current between the gate and the nanotube is used to
monitor the transverse nanotube oscillations.  The characteristic
oscillation frequency is about 1~GHz for a 100~nm
nanotube\cite{Young}, which makes it possible to achieve the
quantum regime at about 50~mK. In Fig.~1(b), a short stiff
nanotube in the STM mode\cite{lieber_AFM} is used to perform
vibrational spectroscopy of an adsorbate loosely bound to a metal
surface.  The position of the atom/molecule modulates the tunnel
current. In the last two examples a kink in the nanotube formed
either mechanically or due to a 5-7 defect plays the role of the
tunnel contact\cite{dekker,kf}. The presence of the defect will
lead to formation of a localized optical phonon mode above the
nano-tube phonon band (above 200~meV\cite{phonons}), that will
couple to the tunnel current by modifying the tunneling matrix
element. Alternatively, the chemical kink defect can be
functionalized by adsorbing an atom or molecule\cite{lieber_func},
whose vibrations will also modify the tunneling between the two
Luttinger legs.

{\it Tunneling Between Two Luttinger Liquids---} We consider the
problem of tunneling between two Luttinger liquids, when the
tunneling is coupled to an external system, such as a quantum
oscillator, or a spin. The Hamiltonian is \be {\cal H}=\Omega T+
{\cal H}^L_1 + {\cal H}^L_2 + {\cal H}_0, \ee where ${\cal H}_0$
is the Hamiltonian for the measured system, referred to as an
oscillator, ${\cal H}^{L}_{1,2}$ are the Luttinger liquid
Hamiltonians for the two leads, with Luttinger parameter $g$ and
with a potential difference $\mu=qV$.  We define the electron
tunneling operator
$T(t)=\Psi^{\dagger}_1(x=0,t)\Psi_2(x=0,t)+h.c.$, where
$\Psi_{1,2}(x,t)$ are fermion operators in the leads. The term
$\Omega$ includes $c$-number terms as well as operators that do
not commute with ${\cal H}_0$.

Following Kane and Fisher\cite{kf}, we consider tunneling via a weak link,
that is tunneling between two semi-infinite leads with $x=0$ located at
the ends of the leads.  We consider the case of repulsive interactions,
$g<1$.  For spinless fermions, the end density of states of a lead at
energy $\epsilon$ is proportional to $\epsilon^{\alpha_{\rm end}}$, with
$\alpha_{\rm end}(g)=1/g-1$.  For carbon nanotubes, where the fermions
have spin but the interactions are only in the charge sector,
one finds $\alpha_{\rm end}(g)=(1/g-1)/4$\cite{kbf}.
Then, the exponent $\alpha$ for a tunnel junction between two leads with
$g_1,g_2$ is $\alpha=\alpha_{\rm end}(g_1)+\alpha_{\rm end}(g_2)$.
For tunneling between two infinite leads, the end density of states is
replaced with $\alpha_{\rm tun}=(g+g^{-1}-2)/2$ for spinless fermions and
$\alpha_{\rm tun}=(g+g^{-1}-2)/8$ for nanotubes.

\begin{figure}[htbp]
\begin{center}
\includegraphics[width = 3.1 in]{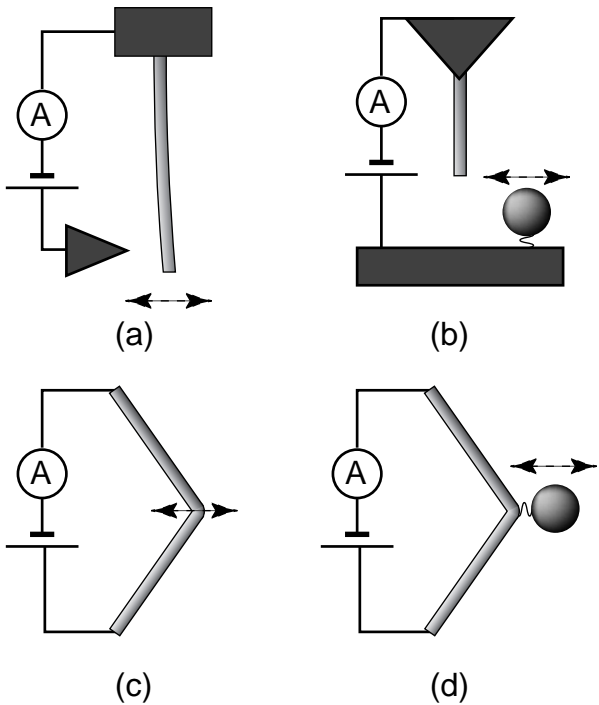}
\vspace{0.0cm} \caption{Experimental realizations of the model.} \label{fig:setup}
\end{center}
\end{figure}

We use a Keldysh formalism\cite{keldysh}.  Our procedure closely
follows that used to study noise in Luttinger liquid
tunneling\cite{chamon}. Let us suppose that initially the
oscillator and leads are decoupled, with density matrices $\rho_0$
for the oscillator and $\rho^L_{1,2}$ for the leads, so that the
full density matrix $\rho=\rho_0 \otimes \rho^L_1 \otimes
\rho^L_2$.  For now, we assume that the leads are at zero
temperature. After the interaction between systems is turned on at
time $t=-\infty$, the systems become coupled. Define the
scattering operator ${\cal S}$ by
\be
\label{kf}
{\cal S}=T_c \, e^{-i\int\limits_{\infty}^{-\infty}
{\cal H} {\rm d}t} e^{-i\int\limits_{-\infty}^{\infty} {\cal H}
{\rm d}t}\equiv 1,
\ee
where the operator $T_c$ denotes time ordering along the Keldysh contours.
Points along the forward branch ($-\infty\rightarrow\infty$)
are ordered with increasing times, while those in the return
($\infty\rightarrow -\infty$) are ordered with decreasing times, with
those in the return branch ordered after those in the forward branch.
We will occasionally use a superscript $f,r$ on $t$ to indicate to
which contour $t$ belongs.
The expectation value of any product of operators
$O(t_1),O(t_2),...$ can be obtained by
\be
\label{ex} \langle O(t_1) O(t_2) ... \rangle= {\rm Tr}\Bigl\{ \rho
\Bigl( T_c \, O(t_1^f) O(t_2^f) ... {\cal S} \Bigr)\Bigr\},
\ee
where we work in the Schroedinger representation throughout.

Tracing out the Luttinger liquids, we obtain
the new scattering operator
\begin{eqnarray}
\label{action}
{\cal S}_{\rm eff}=T_c \,
e^{-i\int\limits_{\infty}^{-\infty} {\cal H}_0 {\rm d}t}
e^{-i\int\limits_{-\infty}^{\infty} {\cal H}_0 {\rm d}t}\times \\
\nonumber e^{\pm \frac{1}{2}\int\limits_{-\infty}^{\infty} {\rm
d}t_1 {\rm d}t_2\, \Omega(t_1)\Omega(t_2)
\frac{2\cos{(\mu(t_1-t_2))}}{(\epsilon\pm i(t_2-t_1))^{\alpha+2}}+
...},
\end{eqnarray}
where the plus sign is chosen before the integral over $t_1,t_2$
if $t_1,t_2$ are in different branches and the minus sign is
chosen if they are in the same branch.  The $\pm$ sign in the
denominator of the exponential is taken positive if $t_2$ is after
$t_1$ and negative if $t_2$ is before $t_1$. We have used $\langle
T(t_1) T(t_2) \rangle_L= 2\cos{(\mu(t_1-t_2))}/(\epsilon\pm
i(t_2-t_1))^{\alpha+2}$, where the expectation value $\langle
\rangle_L$ for the leads is the expectation value for decoupled
Luttinger liquids at zero temperature\cite{chamon}. In this
expectation value, there is an additional factor dependent on the
density of states, which may be absorbed into the normalization of
$\Omega$.  The $...$ in Eq.~(\ref{action}) denote terms higher
order in $\Omega$.  Assuming $\Omega^2 \mu^{\alpha}<<1$, these
higher order terms may be neglected by a power counting, valid for
$\alpha>0$; for $\alpha=0$, there may be logarithmic infrared
divergences which upset this naive power counting, and in this
case we must also assume a sufficiently large voltage to neglect
these terms.

The expectation value of an operator
$O(t)$ becomes
\be
\label{ex2} \langle O(t)\rangle= {\rm Tr}\Bigl\{ \rho_0 \Bigl( T_c
\, \tilde O(t) {\cal S}_{\rm eff} \Bigr)\Bigr\}=\langle \tilde
O(t) \rangle,
\ee
where we have defined $\tilde O=$
\be
\label{otild} \langle O(t_1^f)\rangle_L\mp \int {\rm d}t i
\Omega(t) \langle O(t_1^f) T \rangle_L+...\quad .
\ee
Here the $...$ denote connected expectation values which are higher order
in $\Omega$, and where $\mp$ is chosen negative for $t$ on the forward
contour and positive for $t$ on the return contour.
The operator $\tilde O$ depends only on the oscillator coordinates and not
on the leads.

The exponential in Eq.~(\ref{action})
involves a product $\Omega(t_1)\Omega(t_2)$, where
$t_1,t_2$ may be on either the forward or return contour.  We
introduce $\Omega^s(t_1)=\Omega(t_1^f)+\Omega(t_1^r),
\Omega^a(t_1)=\Omega(t_1^f)-\Omega(t_1^r)$.  Then,
$\pm
\Omega(t_1)\Omega(t_2)
\frac{2\cos{(\mu(t_1-t_2))}}{(\epsilon\pm i(t_2-t_1))^{\alpha+2}}=$
\begin{eqnarray}
\label{as}
-\Omega^a(t_1)\Omega^a(t_2)
{\rm Re}\Bigl(\frac{2\cos{(\mu(t_1-t_2))}}{(\epsilon+i|t_2-t_1|)^{\alpha+2}}
\Bigr)
\\ \nonumber
-2\Omega^a(t_1)\Omega^s(t_2)\theta(t_1-t_2)
{\rm Im}\Bigl(\frac{2\cos{(\mu(t_1-t_2))}}{(\epsilon+i|t_2-t_1|)^{\alpha+2}}
\Bigr)
.
\end{eqnarray}

Again assuming that $\Omega^2 \mu^{\alpha}<<1$, we can make a
further simplification in the exponential of Eq.~(\ref{action}),
by using the Bloch-Redfield approximation\cite{br}, that
$\Omega^{a,s}(t_2)= e^{i{\cal H}_0(t_2-t_1)} \Omega^{a,s}(t_1)
e^{-i{\cal H}_0(t_2-t_1)}$. Corrections to the Bloch-Redfield
approximation arise if operators $\Omega$ are inserted between
$t_1,t_2$; such corrections to ${\cal S}_{\rm eff}$ will be of
order $|t_2-t_1|\Omega^2 \mu^{\alpha+1}$. Let us write
$\Omega=\Omega_{ij}$, where $i,j$ denote eigenstates of ${\cal
H}_0$ with energies $E_{i,j}$. Then, integrating over $t_2$,
\begin{eqnarray}
\label{action2} {\cal S}_{\rm eff}= T_c \,
e^{-i\int\limits_{\infty}^{-\infty} {\cal H}_0 {\rm d}t}
e^{-i\int\limits_{-\infty}^{\infty} {\cal H}_0 {\rm d}t}\times \\
\nonumber e^{\frac{1}{2}\int\limits_{-\infty}^{\infty}{\rm d}t\,
\Bigl(\Omega^{a}(t)\Omega_{ij}^{s}(t)A(\mu,E_i-E_j)-
\Omega^{a}(t)\Omega_{ij}^{a}(t) S(\mu,E_i-E_j)\Bigr) },
\end{eqnarray}
where we define
$S(\mu,\Delta E)=
\int_{-\infty}^{\infty}
2\cos{(\mu t)}e^{i \Delta E t} {\rm Re}\bigl[(\epsilon+i|t|)^{-\alpha-2}\bigr]
{\rm d}t$,
$A(\mu,\Delta E)=
-2 \int_{-\infty}^{0} 2\cos{(\mu t)}e^{i \Delta E t} {\rm
Im}\bigl[(\epsilon+i|t|)^{-\alpha-2}\bigr] {\rm d}t$. The terms in
$\Omega^a\Omega^a$ in Eq.~(\ref{action2}) produce decoherence, and are
equivalent to averaging over a randomly fluctuating field coupled to $\Omega$,
while the terms $\Omega^a\Omega^s$ produce dissipation.

Taking $|E_i-E_j|<\mu$, so that the correct poles in the integrals for $A,S$
are determined by the sign of $\mu$, one finds that
\begin{eqnarray}
\label{asc}
S(\mu,\Delta E)=\frac{1}{2}\Bigl( I(\mu+\Delta E)+I(\mu-\Delta E) \Bigr)
\\ \nonumber
A(\mu,\Delta E)=\frac{1}{2}\Bigl( I(\mu+\Delta E)-I(\mu-\Delta E) \Bigr)+
...,
\end{eqnarray}
where $...$ denotes imaginary terms, possibly singular as $\epsilon\rightarrow 0$, which may be
absorbed into a renormalization of ${\cal H}_0$, and hence dropped. We have defined $I(\mu)=2 \pi
\mu^{\alpha+1}/\Gamma(\alpha+2) $. Eqs.~(\ref{action2},\ref{asc}) are the main results.

{\it Average Current and Noise---} Here $qI(\mu)$ is equal to the
current\cite{wen} flowing at $\Omega=1$. We now recompute the
current within the present formalism, in order to obtain
corrections to the current due to fluctuations in $\Omega$.  The
current operator at time $t^f=0$ is
$J(0^f)=qi\Omega(0^f)(\Psi^{\dagger}_1(0,0^f)\Psi_2(0,0^f)-h.c.)$.
From Eqs.~(\ref{ex2},\ref{otild}), the leading contribution to
$\langle J \rangle$ is of order $\Omega^2$, $\pm i
q\int_{-\infty}^{\infty}{\rm d}t \, \langle \Omega(t) \Omega(0^f)
\rangle 2\sin{(\mu t)} /(\epsilon\pm i t)^{\alpha+2} $. This
vanishes when integrated over $t$ on the forward contour, so we
can assume that $t$ is on the reverse contour. Applying the same
Bloch-Redfield approximation, we get $
q\int_{-\infty}^{\infty}{\rm d}t e^{i(E_i-E_j)t}\langle \Omega(t)
\Omega(0^f) \rangle 2\sin{(\mu t)}/(\epsilon+i t)^{\alpha+2}$.
Doing this integral yields \be \langle J \rangle = q\langle
\Omega_{ij}(0) \Omega(0) \rangle \frac{2\pi}{\Gamma(\alpha+2)}
(\mu+E_i-E_j)^{\alpha+1}.
\ee

We now consider fluctuations in the current, $\langle J(t_1^f) J(t_2^f)
\rangle$.  The order $\Omega^2$ contribution to the current-current correlation
function, from Eq.~(\ref{ex2}), is given by
\begin{eqnarray}
\label{s0} q^2\langle \Omega(t_1^f)\Omega(t_2^f)\rangle \frac{2\cos(\mu (t_2-t_1))}
{(\epsilon+i|t_2-t_1|)^{2+\alpha}},
\end{eqnarray}
This represents the shot noise in the tunnel junction slightly modulated by
oscillator.  To next order in $\Omega^2$, from Eq.~(\ref{ex2}), we must compute
\begin{eqnarray}
\label{manyterms}
-\pm \pm\frac{q^2}{2}\int\limits_{-\infty}^{\infty}
{\rm d}t_3 \, {\rm d}t_4 \,
\Omega(t_1^f) \Omega(t_2^f) \Omega(t_3) \Omega(t_4)
\rangle \times \\ \nonumber
\langle
T(t_1^f) T(t_2^f) T(t_3) T(t_4)
\rangle_L.
\end{eqnarray}
Even for $c$-number $\Omega$, the calculation of Eq.~(\ref{manyterms}) is
involved\cite{chamon}; in this case, the calculation yields a result of order
$\mu^{2\alpha}\Omega^4|t_2-t_1|^{-2}$, plus terms with lower powers of $\mu$.
However, for operator $\Omega$, there is a contribution from
Eq.~(\ref{manyterms}) which is of order $\mu^{2\alpha+2}$, which will dominate
over the previous contribution for $|t_2-t_1|>>\mu^{-1}$, the time regime we
now consider.

The expectation value $\langle T(t_1^f)T(t_2^f)T(t_3)T(t_4)\rangle_L
=\langle T(t_1^f) T(t_3) \rangle_L
\langle T(t_2^f) T(t_4) \rangle_L + t_3\leftrightarrow t_4
+connected$.
The last term, a connected expectation value of $4$ $T$-operators, gives
the contribution of order $\mu^{2\alpha}
\Omega^4 |t_2-t_1|^{-2}$ to Eq.~(\ref{manyterms}).
We ignore this, and consider only the other terms.
Integrating over $t_3,t_4$, and applying a similar procedure to that
used to calculate the current above, we arrive at the following
contribution to $\langle J(t_1) J(t_2) \rangle$:
\begin{eqnarray}
\label{modulate}
&q^2
(\frac{2\pi}{\Gamma(\alpha+2)})^2
(\mu+E_i-E_j)^{\alpha+1}
(\mu+E_j-E_k)^{\alpha+1}\\\nonumber
&\langle
\Omega_{ij}(t_1^r) \Omega_{jk}(t_2^r) \Omega(t_2^f)
\Omega(t_1^f)
\rangle.
\end{eqnarray}
For the expectation value $\langle T(t_1^f) T(t_3) \rangle_L
\langle T(t_2^f) T(t_4) \rangle_L$ to be significant, $t_1-t_3\sim
\mu^{-1}$, $t_2-t_4\sim \mu^{-1}$, justifying the application of
the Bloch-Redfield approximation.  For $|t_2-t_1|>>\mu^{-1}$, this
approximation does not change the time ordering of $t_3,t_4$. As
claimed, Eq.~(\ref{modulate}) is of order $\mu^{2 \alpha+2}$,
reflecting a modulation of the current\cite{korotkov} by the
oscillator.  Eq.~(\ref{modulate}) decays on a time scale of order
the inverse damping coefficient, $1/\gamma$, of the oscillator. In
the case of $c$-number $\Omega$, this term is neglected: when
computing $\langle J J \rangle-\langle J\rangle \langle J
\rangle$, it cancels.

{\it Density Matrix---} For $\Delta E << \mu$, it is possible to write the results above in a
compact form.  Define $\rho(t)$ to be the density matrix for the oscillator at given time $t$.
Then use Eqs.~(\ref{ex2},\ref{action2}) to compute $\langle \dot{\tilde O}(t^f) \rangle$ for any
operator $\tilde O$; the result is be a linear differential equation for the expectation values.
Then use $\langle \dot{\tilde O}(t^f) \rangle \equiv {\rm Tr}(\dot\rho(t) \tilde O)$ to derive the
equation for the density matrix: \be \label{diss} \dot\rho=-i\Bigl[{\cal H}_0,\rho\Bigr]+
\frac{1}{2} \frac{{\rm d}I}{{\rm d}\mu}\Bigl[\Omega,\Bigl\{\Lambda,\rho\Bigr\}\Bigr]-
\frac{I}{2}\Bigl[\Omega,\Bigl[\Omega,\rho\Bigr]\Bigr], \ee with $\Lambda=[{\cal H}_0,\Omega]$.
Comparing to the results for Fermi liquid leads\cite{dima}, the effective temperature
of the oscillator, determined by the ratio of the third (decoherence) term to the second
(dissipation) term in Eq.~(\ref{diss}), is \be T_{\rm eff}=(\alpha+1)^{-1}qV/2. \ee For the
current, we find $\langle J \rangle = q\langle \Omega^2 \rangle I(\mu) +q\langle \Lambda \Omega
\rangle {\rm d}I/{\rm d}\mu$.


{\it Spectral Representation---} While these results were derived for Luttinger liquid leads, they
are more general.  Assuming sufficiently small $\Omega$, we find Eq.~(\ref{action}) with
$2\cos{(\mu(t_1-t_2))}/(\epsilon\pm i(t_2-t_1))^{\alpha+2}$ replaced by the appropriate
expectation value in the leads, $\langle T(t_1) T(t_2)\rangle_L$.  Let the density of states
(particle and hole, respectively) at energy $E$ be $\rho^{p,h}_{1,2}(E)$ in leads $1,2$
respectively. Then, defining $\rho^p(E)=\int {\rm d}E_1 \, \rho^p_1(E_1) \rho^h_2(E-E_1),$
$\rho^h(E)=\int {\rm d}E_1 \, \rho^h_1(E_1) \rho^p_2(E-E_1)$, the expectation value is $\int {\rm
d}E\, ( \rho^p(E) e^{\mp i(E+\mu)(t_1-t_2)}+ \rho^h(E) e^{\mp i(E-\mu)(t_1-t_2)}),$ with the minus
sign chosen if $t_1$ is after $t_2$ and the plus sign otherwise. Then going through the same steps,
we find Eq.~(\ref{action2},\ref{asc}) in general, with $I(\mu)$ replaced by the appropriate
current-voltage characteristic of the device: $I(\mu)=2\pi(\rho^p(-\mu)+\rho^h(\mu))$. The
relation between $S,A$ and $I$ generalizes the fluctuation-dissipation relation derived between
noise and current\cite{fd}. For $\Delta<<\mu$, we arrive at Eq.~(\ref{dvdi}).

{\it Applications---} We now consider the specific case of a harmonic oscillator with frequency
$\omega_0$ and mass $m$, linearly coupled to the tunneling, $\Omega = \Omega_0 + cx$.  We consider a
regime for which the $\Omega_0$ term dominates and the oscillator coordinate $x$ only weakly
modulates the current. For $\mu>\omega_0$, the average current is
\be
J(\mu)=q[I\Omega_0^2 + \frac{c^2 S^2 }{2 m \omega_0A}-\frac{c^2
A}{2 m\omega_0}],
\ee
where $I=I(\mu);S=S(\mu,\omega_0);A=A(\mu,\omega_0)$.

The noise spectrum can be evaluated from
Eqs.~(\ref{s0},\ref{modulate}).  Fourier transforming
Eq.~(\ref{s0}) gives
 the shot noise
contribution
\be
\label{noise} q^2\int dt e^{i\omega
t}\langle\Omega(0)\Omega(t)\rangle_L\langle T(0)T(t)\rangle_L =
2q^2\Omega_0^2 I,
\ee
for $|\omega|\ll|\mu|$.  The modulation,
Eq.~(\ref{manyterms}), to leading order in $c^2$
from Eq.~(\ref{action2}),
yields
\be
\label{signal} \frac{q^2 \Omega_0^2 c^4 } {m^2} \frac{(I+S)^2
S-A^2(S+2I)} {(\omega^2-\omega_0^2)^2+\gamma^2\omega^2},
\ee
where $\gamma = c^2 A(\mu,\omega_0)/(m\omega_0)$.  At the peak,
the signal-to-noise ratio, Eq.~(\ref{signal}) divided by
Eq.~(\ref{noise}), is approximately $2I^2/A^2$.

{\it Acknowledgements---} This work was supported by DOE
W-7405-ENG-36, NSF DMR-0121146 (DM), and DARPA MOSAIC (IM). We
thank M. Paalanen for encouragement, and K. Schwab and A. Shnirman
for useful discussions.


\begin{thebibliography}{99}

\bibitem{naka} Y. Nakamura, Y.A. Pashkin, and J.S. Tsai, Nature (London) {\bf
398}, 786 (1999).

\bibitem{saclay} D. Vion, {\em et al.},  Science {\bf 296}, 886
(2002).

\bibitem{kansas} Y. Yu, {\em et al.}, Science, {\bf 296}, 889 (2002).

\bibitem{mandelstam} V.B. Braginsky and F.Ya. Khalili, {\em Quantum
Measurement}, (Universiy Press, Cambridge 1992).


\bibitem{gurvitz} S. A. Gurvitz, Phys. Rev. B {\bf 56}, 15215 (1997); S. A. Gurvitz and Ya. S. Prager
Phys. Rev. B {\bf 53}, 15932 (1996).

\bibitem{shnirRMP}  Y. Makhlin, G. Schon, and A. Shnirman, Rev. Mod.
Phys. {\bf 73}, 357 (2001).

\bibitem{korotkov} A. N. Korotkov, Phys. Rev. B {\bf 60}, 5737 (1999); A. N. Korotkov and D. V.
Averin, Phys. Rev. B {\bf 64}, 165310 (2001).

\bibitem{goran} G. Johansson, A. K\"{a}ck, and G. Wendin, Phys. Rev. Lett.
{\bf 88}, 046802 (2002).

\bibitem{dima} D. Mozyrsky and I. Martin, Phys. Rev. Lett. {\bf 89}, 018301 (2002).

\bibitem{lutt} D. Mattis and E. Leib, J. Math. Phys {\bf 15}, 609 (1965).

\bibitem{dekker} Z. Yao, H.W.Ch. Postma, L. Balents, and C. Dekker,
Nature (London) {\bf 402}, 273 (1999).

\bibitem{2c} W. Apel and T. M. Rice, Phys. Rev. B {\bf 26}, 7063 (1982).

\bibitem{2cw} D. L. Maslov and M. Stone, Phys. Rev. B {\bf 52}, 5539 (1995).

\bibitem{Young} M.M.J. Treacy, T.W. Ebbesen, and J.M. Gibson, Nature (London)
{\bf 381}, 678 (1996); R.S. Ruoff and D.C. Lorents, Carbon {\bf 33}, 925
(1995).

\bibitem{lieber_AFM} J.H. Hafner, C.L. Cheung, and C.M. Leiber, Nature (London)
{\bf 398}, 761 (1999).

\bibitem{kf} C. L. Kane and M. P. A. Fisher, Phys. Rev. Lett. {\bf 68},
1220 (1992).

\bibitem{phonons} D. S\'anchez-Portal, {\em et al.}, Phys. Rev. B {\bf 59},
12678 (1999).

\bibitem{lieber_func} S.S. Wong, {\em et al.}, Nature (London) {\bf 394}, 52
(1998).

\bibitem{keldysh} L. V. Keldysh, Soviet Phys. JETP {\bf 20}, 1018 (1965).

\bibitem{chamon} C. D. Chamon, D. E. Freed, and X. G. Wen,
Phys. Rev. B {\bf 51}, 2363 (1995).

\bibitem{kbf} C. Kane, L. Balents, and M. P. A. Fisher, Phys. Rev. Lett.
{\bf 79}, 5086 (1997).

\bibitem{br} F. Bloch, Phys. Rev. {\bf 70}, 460 (1946); A.G. Redfield, IBM J.
Res. Dev. {\bf 1}, 19 (1957).

\bibitem{wen} X. G. Wen, Phys. Rev. B {\bf 44}, 5708 (1991).

\bibitem{fd} D. Rogovin and D.J. Scalapino, Ann. Phys. (N.Y.) {\bf 86}, 1
(1974).

\end{thebibliography}
\end{document}